\documentclass{amsart}
\usepackage{amssymb,latexsym,color}

\usepackage{multicol, color}

\newcounter{mnote}
\newcounter{mnoteE}
\usepackage{graphicx,psfrag}
\numberwithin{equation}{section}

\def\cA{\mathcal{A}}

\def\cP{P}
\def\cD{\mathcal{D}}
\def\cO{\mathcal{O}}
\def\cG{\mathcal{G}}
\def\cK{\mathcal{K}}

\def\k0{\kappa_0}

\def\kfup{\overline{\kappa}}

\def\ktau{\kappa_{\text{T}}}

\def\lgl{\langle}
\def\rgl{\rangle}
\def\bfe{{\mathsf{e}}}
\def\uE{\underline{\mathsf{E}}}
\def\ue{\underline{\mathsf{e}}}

\def\bfe{{\mathsf{e}}}
\def\bfoE{{\mathsf{E}}}
\def\bfoP{{\mathsf{P}}}
\def\bfE{{\mathcal{E}}}
\def\bfP{{\mathcal{P}}}
\def\bfF{{\bf{P}}}

\def\bR{{\mathbb R}}

\def\lgl{\langle}
\def\rgl{\rangle}

\def\curl{\text{curl} \ }
\def\ve{\varepsilon}
\def\om{\omega}

\def\OOO{\mathcal{O}}

\begin{document}
\title[Vorticity coherence effect on energy-enstrophy bounds ]
{
Effect of vorticity coherence on energy-enstrophy bounds for the 3D
Navier-Stokes equations}
\author{R. Dascaliuc$^{1}$}\address{$^1$Department of Mathematics\\
Oregon State University\\ Corvallis, OR 97331}
\author{Z. Gruji\'c$^2$}
\address{$^2$Department of Mathematics\\
University of Virginia\\ Charlottesville, VA 22904}
\author{M. S. Jolly$^3$}
\address{$^3$Department of Mathematics\\
Indiana University\\ Bloomington, IN 47405}
\email[R. Dascaliuc]{dascalir@math.oregonstate.edu}
\email[Z. Grujic]{zg7c@virginia.edu}
\email[M. S. Jolly]{msjolly@indiana.edu}

\thanks{R.D.~was supported in part by National Science Foundation grant DMS-1211413.  Z.G.~acknowledges support of the Research Council of Norway
via grant 213474/F20, and the NSF via grant DMS 1212023.  M.S.J.~was supported in part by NSF grant numbers DMS -1008661 and DMS-1109638.}

\date{\today}

\begin{abstract}
Bounding curves in the energy,enstrophy-plane are derived for the 3D Navier-Stokes equations
under an assumption on coherence of the vorticity direction.  The analysis in the critical case where the
direction is H\"older continuous with exponent $r=1/2$ results in a curve with extraordinarily large maximal enstrophy (exponential in Grashof),
in marked contrast to the subcritical case, $r>1/2$ (algebraic in Grashof).
 \end{abstract}
\maketitle

\begin{center}
{\it Dedicated to Ciprian Foias in appreciation of his unbounded generosity}
\end{center}

\section{Introduction}\label{sec1}
Numerical simulations of turbulent flows (cf. \cite{AKKG87, JWSR93, SJO91, VM94}) reveal 
regions of \emph{intense vorticity} dominated by coherent vortex structures; more
specifically, \emph{vortex filaments}. 
One of the imminent morphological signatures of the filamentary geometry is the \emph{local coherence of the vorticity direction}; it turns out
that this property of turbulent flows leads to the \emph{geometric depletion 
of the nonlinearity}. 

The pioneering work in this direction was made by Constantin. 
He obtained in \cite{C94} a singular
integral representation of the stretching factor in the evolution of the vorticity magnitude
featuring a geometric kernel that is depleted by local coherence of the vorticity direction,
a purely geometric condition. This led to the first rigorous confirmation in \cite{CF93} of the 
local anisotropic
dissipation in the 3D NSE: a theorem stating that as long as the vorticity
direction is Lipschitz-coherent (in the regions of high vorticity), the $L^2$-norm of the vorticity
is controlled, and no finite time blow-up can occur. 

Subsequent work delved further into this geometric condition.  The relaxation of the Lipschitz-coherence condition
to a $\frac{1}{2}$-H\"older condition was made in \cite{daVeigaBe02}, followed
by a full \emph{spatiotemporal 
localization} of the $\frac{1}{2}$-H\"older condition in \cite{GrGu10-1}. A family of local, 
hybrid, geometric-analytic regularity criteria including a \emph{scaling invariant} improvement 
of the $\frac{1}{2}$-H\"older condition was presented in \cite{Gr09}. 
Studies of coherence of the vorticity direction-type regularity
criteria on bounded domains in the cases of no-stress and no-slip
boundary conditions were presented in \cite{daVeigaBe09} and \cite{Beirao}, respectively.

In this paper we derive a bounding curve for the weak attractor in the plane spanned by
energy $\bfe$ and enstrophy $\bfE$ under the $\frac{1}{2}$-H\"older condition in \cite{daVeigaBe02}.
These curves suggest two scenarios in which the Taylor wave number
$\ktau$ can satisfy
\begin{align} \label{ktau}
\ktau:=\left(\frac{\lgl \bfE\rgl}{\lgl \bfe\rgl}\right)^{1/2} \gg \kfup
\end{align}
where $\kfup$ is the largest wavenumber in the force.  One is where both energy and enstrophy are small; 
the other is where there are brief, intense bursts of large enstrophy.  
It is shown in \cite{FMRT2,FJMRT} that \eqref{ktau} guarantees 
the energy cascade predicted by the Kolmogorov theory of 
turbulence \cite{K41}.  The $\frac{1}{2}$-H\"older condition is essential for our analysis, of course.
Yet, if the resulting curve turns out to be sharp, this condition still allows for 
excursions with extraordinarily large enstrophy; the maximum value achieved by this curve 
is $\bfE_{\text{max}}=\cO(\exp(G^2))$, where 
$G$ is the Grashof number.  This would mean that if the $\frac{1}{2}$-H\"older condition
were valid, solutions that are in fact regular might appear to blow-up in finite time under numerical simulation.

The approach for the bounding curves follows that of Foias and Prodi (see pages 59-61 in \cite{Fstat}
as well as \cite{FoiasGuillope,DFJ}).
Each curve solves an ordinary differential equation in which 
$\bfe$, $\bfE$ are the independent and dependent variables respectively.  The ODE
is formed by the quotient of estimates for the growth rates $d\bfe/dt$ and $d\bfE/dt$ in a region of the
$\bfe,\bfE$-plane where $d\bfe/dt < 0$.  This technique was applied to the 2D NSE in both the energy,enstrophy-plane in \cite{DFJ}, as well as in the enstrophy,palinstrophy-plane in \cite{DFJ4}, the latter being more relevant 
to the Kraichnan theory of 2D turbulence \cite{K67}.  Curves which close were also found in this manner in \cite{FJL} for the 3D Leray-$\alpha$ and the 3D Navier-Stokes-$\alpha$ models, whose solutions are global in time and approach those
of the 3D NSE as the filter width parameter $\alpha \to 0$ \cite{CTV02,VTC07}.  As expected, however, these bounding curves for the $\alpha$-models blow up as $\alpha \to 0$.

For comparison, we include in this paper curves that somewhat limit solutions for the full NSE, i.e., without any conditions.  Of course,  using current
techniques for estimating growth rates, those
curves do not close.  In \cite{Doering} Doering uses the system of differential inequalities given by growth rate 
estimates to show that for initial data satisfying $\bfE_0 \lesssim \nu^4/\bfe_0$ the enstrophy remains bounded.  The maximal enstrophy growth rate for given enstrophy was studied numerically in \cite{LuDoering}.

In order to properly assess the significance of the $\frac{1}{2}$-H\"older condition
in the realm of dynamics within the energy, enstrophy-plane, we also include the results describing 
the effects
of the sub-critical coherence case where the H\"older exponent satisfies $1/2 < r \le 1$, as well as
the effects of an example of the more traditional scaling-invariant conditions on the vorticity
magnitude, namely, the small, uniform-in-time, $L^\frac{3}{2}$-spatial integrability condition.
It is worth observing that there is a \emph{qualitative jump} between the maximal enstrophy allowed 
in the sub-critical and in the $\frac{1}{2}$-H\"older coherence scenarios; more precisely, from algebraic 
in $G$ to exponential in $G$. Perhaps more intriguing is the comparison between the whole
range of the H\"older coherence conditions and the $L^\frac{3}{2}$-condition. With respect to
the scaling inherent to the 3D NSE, the $L^\frac{3}{2}$-condition is scaling invariant, and the
H\"older coherence conditions are all sub-critical; hence--in this metric--the 
$L^\frac{3}{2}$-condition is a weaker condition. In contrast, the $L^\frac{3}{2}$-condition
produces the most restrictive maximal curve in the energy, enstophy-plane, and is--in this sense--a
\emph{dynamically stronger condition}.

After some preliminaries in Section \ref{sec2}, we derive the bounding curves for the full
NSE in Section \ref{full}.  The precise formulation
of the H\"older condition is given in Section \ref{deplete}.  The critical case,  
H\"older exponent $r=1/2$, is treated in Section \ref{critic} resulting in a bounding curve
in terms of incomplete gamma functions, followed by sharp estimates for the maximum enstrophy value for this curve.
Section \ref{subcritsec} and Section \ref{scaling-invariant} address the sub-critical H\"older coherence
and the scaling-invariant $L^\frac{3}{2}$-spatial integrability cases, respectively.


\section{Preliminaries}\label{sec2}
We consider the incompressible Navier-Stokes equations (NSE) in a bounded, connected, open set
$\Omega \in \bR^3$
\begin{align}\label{NSE}
\frac{\partial u}{\partial t} -\nu\Delta u + (u\cdot\nabla)u + \nabla p
&= F  \;, \quad x \in \Omega
\\ 
\label{incomp}
\nabla\cdot v &= 0\;,
\end{align}
with either periodic or no-slip boundary conditions.
The NSE can be written in functional form as 
\begin{align}\label{fNSE}
\frac{d u}{d t} + \nu A u + B(u,u) =f\;, \quad u \in H
\end{align}
where $H$ is an appropriate ($L^2$-based) Hilbert space, $A=-\cP\Delta$ is the Stokes operator, $B(u,v)=\cP(u\cdot\nabla)v$, $f=\cP F$, 
and $\cP$ is the Helmholtz-Leray projector onto divergence-free functions.  In this case, $A$ is a positive definite self-adjoint operator with compact inverse, and we will denote by $\lambda(>0)$ its smallest eigenvalue. To measure the size of the body force $f$, we introduce the dimensionless Grashof number
\[
G= \frac{\|f\|_2}{\nu^2\lambda^{3/4}}\;,
\]
where $\|\cdot\|_2$ is the $L^2$-norm on $\Omega$.
A more detailed description of this functional setting for the NSE can be found in  \cite{T97}.

One of the main results of the (yet incomplete) NSE regularity theory is the existence of {\em Leray-Hopf weak solutions} to the initial value problem associated with (\ref{fNSE}).  For an arbitrary $T>0$ (fixed), any Leray-Hopf
solution is $\cD(A)$-valued except, possibly, on a closed subset of $[0,T]$ of 1/2-dimensional
Hausdorff measure 0.  Thus it is regular on an open set which
can then be written in the worst case as a union of countably many mutually
disjoint open intervals, a set that is dense in $[0,\infty)$.  See \cite{CF88}, \cite{Lemarie02book}, and \cite{TemamNSEBook} for general background on the regularity theory of the NSE.

We will use several well-known properties of the nonlinear term $B$,
namely 
\begin{align}\label{orthog}
B(u,v,v)=0\;,\quad  \forall \ u \in H\;, \ v \in V=\cD(A^{1/2})
\end{align}
and
\begin{align}\label{Bineq}
|(B(u,u),Au)| \le c\|A^{1/2}u\|_2^{3/2}\|Au\|_2^{3/2}\;,
\end{align}
 (see (6.18) and (9.22) in \cite{CF88}).
Throughout the paper $c$
will denote a generic, dimensionless constant, which will change from line to
line, except when distinguished by a subscript.

Within each interval of regularity, we can take the $L^2$-scalar product of \eqref{fNSE} with $u$, and then applying \eqref{orthog},
the Cauchy-Schwarz  and Young inequalities, one has the standard differential relation for energy
\begin{align}\label{energy1}
\frac{d\|u\|^2_2}{dt}+ 2\nu \|A^{1/2}u\|_2^2 \le 2\|f\|_2 \|u\|_2=2\nu^2\lambda^{3/4}G\bfe^{1/2} \le \nu^3\lambda^{1/2}G^2 + \nu\lambda \|u\|^2_2\;,
\end{align}
which, together with the Poincar\'e and Gronwall inequalities, implies that for regular solutions
$$
\limsup_{t\to\infty}\|u\|^2_2 \le \frac{\|f\|_2^2}{\nu^2\lambda^2}=\frac{\nu^2G^2}{\lambda^{1/2}}.
$$
In fact this property holds for general Leray-Hopf solutions.

More precisely, the long-time behavior of (\ref{fNSE}) is encoded in the weak attractor, $\cA_w$ (first introduced in \cite{FTwatt}), the subset of Leray-Hopf solutions that are globally bounded in $H$ --  forward and backward in time.  This set weakly attracts all Leray-Hopf solutions as $t\to\infty$. By using the Leray-Hopf energy inequality one can establish that $\cA_w$ is contained in a ball in $H$ centered at $0$ with radius ${\nu G}/{\lambda^{1/4}}$, i.e.,
\begin{align}\label{w_attr_bd}
\cA_w\subseteq B(0, {\nu G}/{\lambda^{1/4}})\;.
\end{align}
In particular $B(0, {\nu G}/{\lambda^{1/4}})$, is a (forward in time) invariant set for the Leray-Hopf solutions. Therefore, in what follows our study of energy-enstrophy relations will be focused {\em exclusively on this ball}.

We conclude by noting that the above theory of Leray-Hopf solutions will be less relevant in the no-slip case studied in Sections \ref{critic}-\ref{subcritsec}, since the vorticity coherence assumption (\ref{betacond}) combined with (\ref{bdryint}) imply that the solution is in fact regular, and therefore estimates like (\ref{energy1}) will hold for all $t$.

\section{Bounds for the full 3D Navier-Stokes equations} \label{full}

While an overall bound on enstrophy for the 3D NSE remains an open question, the partial regularity results described above will allow us to derive certain curves in the energy,enstrophy-plane which limit the long time behavior of Leray-Hopf solutions, without any additional regularity assumptions.  

In this section we consider periodic boundary conditions on $\Omega=[0,L]^3$,
take the phase space $H$ for \eqref{fNSE} to be the closure in $L^2(\Omega)^3$
of all divergence-free, mean zero trigonometric polynomials,
and denote the energy, enstrophy and palinstrophy by
\begin{align*}
\bfe=\|u\|_2^2\;, \quad \bfoE=\|A^{1/2}u\|_2^2\;, \quad \bfoP=\|Au\|_2^2\;.
\end{align*}


As noted before, we will concentrate on the globally bounded solutions, for which (\ref{w_attr_bd}) holds, i.e:

\begin{align}\label{Kbnd}
\bfe \le \bfe_0=\frac{\|f\|_2^2}{\nu^2\lambda^2}=\frac{\nu^2G^2}{\lambda^{1/2}}\;.
\end{align}
Now on each interval of regularity, take the $L^2$-scalar product of \eqref{fNSE} with $Au$, apply \eqref{Bineq},
and then proceed as in \eqref{energy1} to find
\begin{align*}
\frac{1}{2}\frac{d\bfoE}{dt}+ \nu \bfoP &= \|f\|_2\bfoP^{1/2} + c\bfoE^{3/4}\bfoP^{3/4} \\
&\le \frac{\|f\|_2^2}{\nu}+\frac{\nu \bfoP}{4}+\frac{c\bfoE^3}{\nu^3} + \frac{\nu \bfoP}{4}\;.
\end{align*}
Again by the Cauchy-Schwarz inequality we have $\bfoE^2\le \bfe\bfoP$, so that
\begin{align}
\frac{d\bfoE}{dt} &\le 2\nu^3\lambda^{3/2}G^2+\frac{c_1\bfoE^3}{\nu^3} - \nu \bfoP \nonumber \\
&\le 2\nu^3\lambda^{3/2}G^2 +\frac{c_1\bfoE^3}{\nu^3} - \nu \frac{\bfoE^2}{\bfe} \label{fromhere} \;.
\end{align}
Note that $d\bfoE/dt \le 0$ provided
\begin{align}\label{negcond}
\bfe \le \Psi(\bfoE)= \frac{\nu^4 \bfoE^2}{2\nu^6 \lambda^{3/2}G^2 + c_1\bfoE^3}\;,
\end{align}
and that the nonzero critical number for $\Phi$ is
\begin{align}\label{critno}
\bfoE_{1}=\left(\frac{4}{c_1}\right)^{1/3}\nu^2\lambda^{1/2}G^{2/3}\;.
\end{align}

For some $\eta>1$ consider the region where we have both
\begin{align} \label{parfull}
\bfoE\ge \eta \nu \lambda^{3/4}G\bfe^{1/2}
\end{align}
and
\begin{align*} 
\frac{c_1\bfoE^3}{\nu^3} \ge 2\nu^3\lambda^{3/2}G^2 \;, 
\end{align*}
or equivalently
\begin{align}\label{fullcond22}
\bfoE \ge \uE =\left(\frac{2}{c_1}\right)^{1/3}\nu^2\lambda^{1/2}G^{2/3}=2^{-1/3}\bfoE_1\;.
\end{align}
Using \eqref{parfull} and \eqref{fullcond22}  in \eqref{fromhere}, we have 
\begin{align}\label{top}
\frac{d\bfoE}{dt} \le 2 c_1\nu^{-3}\bfoE^3 - \eta \nu^2\lambda^{3/4} G\bfe^{-1/2}\bfoE \quad \mbox{a.e.}\ t\in[0,\infty)\;,
\end{align}
while \eqref{parfull} in \eqref{energy1} gives
\begin{align}\label{bottom}
\frac{d\bfe}{dt} \le  -2(\eta-1)\nu^2\lambda^{3/4}G\bfe^{1/2}\quad \mbox{a.e.}\ t\in[0,\infty)\;.
\end{align}

To find a bounding curve we proceed as in \cite{Fstat,FoiasGuillope,DFJ}.  It is constructed
as a solution to an ODE in the variables $\bfe$, $\bfoE$, in decreasing $\bfe$.   As long as 
\eqref{parfull} holds so that the upper bound \eqref{bottom} is not positive, and as 
long as the bound in right hand side of \eqref{top} is nonnegative, the slope of
this curve is determined by quotient of the two bounds.
The bounding curve is then the solution to the Bernoulli ODE
\begin{align}
\frac{d\bfoE}{d\bfe} &= \frac{2 c_1\nu^{-3}\bfoE^3 -\eta \nu^2\lambda^{3/4} G\bfe^{-1/2}\bfoE}
{-2(\eta-1)\nu^2\lambda^{3/4}G\bfe^{1/2}} \nonumber \\
&=\frac{1}{2}\frac{\eta}{\eta-1} \frac{\bfoE}{\bfe} - \left[ \frac{c_1}{(\eta-1)\nu^5\lambda^{3/4}G}\right] \frac{\bfoE^3}{\bfe^{1/2}} \label{BernODE}\;,
\end{align}
whose solution can be written as 
\begin{align}\label{Bernie}
\bfoE=\Phi(\bfe)=\left\{\left(\frac{\bfe_0}{\bfe}\right)^\alpha\bfoE_0^{-2} + \beta\left(\bfe^{1/2}-\bfe^{-\alpha} \bfe_0^{1/2+\alpha}\right)\right\}^{-1/2}\;,
\end{align}
where 
$$
\alpha=\frac{\eta}{\eta-1}\quad \text{and} \quad \beta= \frac{4c_1}{(3\eta-1)\nu^5\lambda^{3/4}G}\;.
$$
The function $\Phi$ has a vertical asymptote at $\bfe=\bfe_*$, where 
$$
\bfe_*^{\alpha+1/2}=\bfe_0^{\alpha+1/2}-\frac{\bfe_0^\alpha}{\beta \bfoE_0^2}\;.
$$
We assume that \eqref{fullcond22} holds for initial data $(\bfe_0,\bfoE_0)$ on the parabola 
\begin{align}\label{parfulleq}
\bfoE =\eta \nu \lambda^{3/4}G\bfe^{1/2}\;,
\end{align}
 so that in terms of $\eta$, this asymptote is at 
$$
\bfe_*(\eta)=\bfe_0\left(1-\frac{3\eta-1}{4c_1\eta^2G^4}\right)^{\frac{2(\eta-1)}{3\eta-1}}\;.
$$

We next solve \eqref{BernODE} with initial data $(\bfe_1,\bfoE_1)$ given by the maximum of $\Psi$ in \eqref{negcond}, where $\bfoE_1$ is as in \eqref{critno} and 
$$
\bfe_1=\frac{1}{6}\left(\frac{4}{c_1}\right)^{2/3}\nu^2 \lambda^{-1/2}G^{-2/3}\;.
$$
The solution curve with this initial data would intersect the parabola \eqref{parfulleq}
at the value of $\bfe$ satisfying 
\begin{align}\label{too nasty}
\bfe^{1/2+\alpha} -\gamma \bfe^{\alpha-1} + \delta
  =0
\end{align}
where
$$
\gamma= \frac{1}{\beta\eta^{2}\nu^{2}\lambda^{3/2}G^{2}}\quad \text{and} \quad \delta=\frac{\bfe_1^\alpha}{\beta\bfoE_1^2} - \bfe_1^{1/2+\alpha} \;.
$$
Substituting the values for $\bfe_1$, $\bfoE_1$, one can show  that $\delta>0$ provided 
$$
\eta <  1 + \frac{4c_1}{3\sqrt{6}}\left(\frac{4}{c_1}\right)^{5/6}\;.
$$

\begin{figure}[h]
      \psfrag{G}{{\tiny $\OOO(G^2)$}}
      \psfrag{G2}{{\tiny $\OOO(G^2)$}}
      \psfrag{F}{{\tiny${\bfF}$}}
      \psfrag{E}{{\tiny ${\bfoE}$}}
      \psfrag{G4}{{\tiny $cG^{4}$}}
      \psfrag{parabola}{{\tiny$\bfE=G\sqrt{\bfe}$}}
      \psfrag{e}{{\tiny${\bfe}$}}
      \psfrag{phi}{{\tiny$\bfe=\Psi(\bfoE)$}}
      \psfrag{p}{{\tiny${(\bfe_2,\bfoE_2)}$}}
      \psfrag{pbar}{{\tiny${(\ue,\uE)}$}}
      \psfrag{psi}{{\tiny$\bfe=\Phi(\bfoE)$}}
      \psfrag{G2/3}{{\tiny $\bfoE_1= \OOO(G^{2/3})$}}
      \psfrag{G-23}{{\tiny $\bfe_1=\OOO(G^{-2/3})$}}
      \psfrag{tewards singularity}{{\tiny Towards singularity}}
      \psfrag{recurent region}{{\tiny recurrent region}}
      \psfrag{intermittent}{{\tiny intermittent region}}
      \psfrag{vortform}{{\tiny 50\% of max $|\omega|$}}
      \psfrag{stretchform}{{\tiny 20\% of max $\frac{\om\cdot\nabla u \cdot \omega}{|\om|^2}$}}
      \psfrag{vorticity contour}{{\tiny vorticity contour}}
      \psfrag{stretching factor contour}{{\tiny stretching factor contour}}
      \psfrag{(b)}{{\tiny (b)}}
      \psfrag{(c)}{}
      \psfrag{I}{\tiny I}     
      \psfrag{II}{\tiny II}     
      \psfrag{III}{\tiny III}    
      \psfrag{IV}{\tiny IV}
  \centerline{\includegraphics[scale=.3]{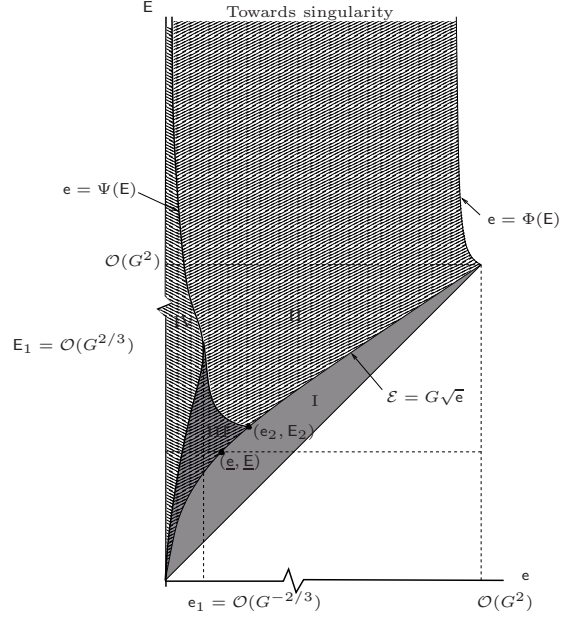}}
\caption{Bounds for the full 3D NSE.  Region I is recurrent.  If the solution blows up, it must enter region II.
If it enters regions III, IV it must enter region I.  In region IV the enstrophy is always decreasing, while
in regions II,III, IV, the energy is always decreasing.}
\label{fig9}
\end{figure}

Dropping the $\gamma$-term in
\eqref{too nasty}, we obtain as a lower bound for its root
$$
\bfe_2>\delta^{\frac{2}{2\alpha+1}}\nu^2\lambda^{-1/2}G^{2/3} \;.
$$
For large $G$, we then have that $\bfe_2 > \ue$,  where $(\ue,\uE)$ is the point
of intersection of the line $\bfoE=\uE$ and the parabola in \eqref{parfulleq}.
A qualitative sketch of the shaded region in which the weak attractor must lie is given in Figure \ref{fig9}.  It is not surprising that this region is not closed, since global regularity is still an open question.   As far as these 
estimates show, the sufficient condition (large $\ktau$, see \eqref{ktau}) for an energy cascade
can be met either by the solution spending most of its time near zero, or with repeated excursions
into the intermittent region.

\section{Vortex stretching depletion}\label{deplete}
We now show how a geometric condition of the direction of vorticity
allows us to close the region for the weak attractor.
For convenience in the analysis to follow, we now consider no-slip boundary conditions
\begin{align}\label{BC}
u(x)&= 0\;, \quad x \in \Gamma=\partial \Omega\;.
\end{align}

Taking the curl of \eqref{NSE} gives the equation for the vorticity $\omega=\curl u$
\begin{align}\label{omegaeqn}
\frac{\partial \omega}{\partial t} -\nu\Delta \omega+ (u\cdot\nabla)\omega 
&= (\omega\cdot\nabla)u + \curl F  \;, \quad x \in \Omega
\end{align}
The evolution of the enstrophy is found by taking the scalar product of \eqref{omegaeqn} with $\omega$
in $L^2(\Omega)$
\begin{align}\label{enstbal}
\frac{1}{2}\frac{d}{dt}\|\om\|_2^2+\nu\|\nabla \om\|_2^2 -\nu \int_{\Gamma}  \frac{\partial \omega}{\partial n} \cdot \omega \ d \Gamma 
=((\om\cdot\nabla)u,\omega)+(\curl F, \om)\;.
\end{align}
The obstacle to proving global existence of solutions is the vortex stretching term $((\om\cdot\nabla)u,\omega)$.  One way to mitigate its effect is to impose a geometric condition on it
in physical space.  


This amounts to a coherence assumption on the direction of the vorticity of the form
\begin{align}\label{betacond}
|\sin\theta(\omega(x),\omega(y))| \le c\lambda^{r/2}|x-y|^r
\end{align}
where
$\theta$ is the angle between the two vectors, and
$|\cdot|$ denotes either $\|\cdot\|_{\bR^3}$, or absolute value.

\section{The critical case}\label{critic}
Following the work of Constantin and Fefferman in \cite{CF93,C94}, H. Beir\~ao da Veiga shows  in
  \cite{Beirao} that condition \eqref{betacond} with $r=1/2$, 
along with the additional assumption
\begin{align}\label{bdryint}
\int_{\Gamma}  \frac{\partial \omega}{\partial n} \cdot \omega \ d \Gamma 
\le \delta\|\nabla \omega\|_2^2 + \lambda \psi(t) \| \omega \|_2^2\end{align}
for some constant $\delta < 1$ and some $\psi \in L^{1}(0,T)$,
is sufficient for regularity of the solution to
the NSE on the interval $[0,T]$.  
To do so he splits the vorticity as 
\begin{align*}
\omega(x)=\omega^{(1)}(x)+\omega^{(2)}(x)=\begin{cases} \omega^{(1)}(x)  
& \text{if } | \omega(x)|\le \mu\lambda\nu \\
\omega^{(2)}(x)  & \text{if }  | \omega(x)| >  \mu\lambda\nu
\end{cases}
\end{align*}
for some $\mu  > 0$.
This leads to an expansion of the trilinear term into eight summands:
\begin{align*}
((\omega\cdot\nabla) u \cdot \omega) (x)&=\sum_{\alpha,\beta,\gamma=1}^2 \cK_{\alpha\beta\gamma}(x) \\
&=-\sum_{\alpha,\beta,\gamma=1}^2 \epsilon_{jkl} \ \omega_i^{(\alpha)}(x)\omega_j^{(\beta)}(x) \int_\Omega \frac{\partial^2 \cG (x,y)}{\partial y_k\partial x_i}\omega_l^{(\gamma)}(y) \ dy
\end{align*}
where 
$$
\epsilon_{jkl}=\begin{cases} 1 & (i,j,k) \text{ even permutation} \\
-1& (i,j,k) \text{ odd permutation} \\
0 & i,j,k \text{ all equal} \end{cases}\;, \quad \text{and}\quad
\cG(x,y)=\frac{1}{4\pi|x-y|}+g(x,y)
$$
for some regular function $g$.   The only summand that uses \eqref{betacond} is
$\cK_{222}$.   Since we will modify slightly
Beir$\tilde{\rm{a}}$o da Veiga's estimate, we recall part of his argument. First he shows that
$$
| \cK_{222}(x)|  \le c \int_\Omega\frac{|\omega^{(2)}(x)|^2|\omega^{(2)}(y)|}{|x-y|^3}|\sin \theta(x,y)|\ dy\;.
$$
Then, writing $I$ for the Riesz potential
$$
I(x)=\int_\Omega |\omega(y)| \frac{dy}{|x-y|^{5/2}}\;,
$$
using \eqref{betacond} with $r=1/2$,  and H\"older's inequality, he finds that
\begin{align}\label{HolderPoint}
\int_\Omega |\cK_{222}(x)| \ dx \le c\lambda^{1/4} \int_\Omega |\omega(x)|^2 I(x) \ dx \le c\lambda^{1/4} \|\omega\|_2\|\omega\|_6 \| I \|_3 \end{align}
As in \cite{Beirao}, we apply the Hardy-Littlewood-Sobolev inequality \cite{Stein} $\| I \|_3 \le c\|\omega\|_2$ and 
the Sobolev embedding $H^1_0(\Omega) \subset L^6(\Omega)$, but then use the Poincar\'e inequality before
applying Young's inequality 
\begin{equation}
\begin{aligned}\label{poincuse}
 \int_\Omega |\cK_{222}(x)| \ dx  &\le  c\lambda^{1/4} \|\omega\|_2(\lambda^{1/2}\|\omega\|_2 + \|\nabla\omega\|_2)  \|\omega\|_2 \\
  &\le c\lambda^{1/4}  \|\nabla\omega\|_2  \|\omega\|_2^2  \\
   &\le \ve \nu \|\nabla \omega\|_2^2 + c \lambda^{1/2}\frac{\|\omega\|_2^4}{\ve\nu}     \;.               
\end{aligned}
\end{equation}
Otherwise, except for the inclusion of dimensional factors involving $\nu$ and $\lambda$,
we treat the bounds for the remaining summands as in \cite{Beirao}.  They amount to the three cases 
\begin{align*} 
|\int_\Omega \cK_{\alpha\beta 2}(x) \ dx| \le c \mu\nu\lambda \|\om\|_2^{2} \;, \qquad \text{for} \ (\alpha,\beta) \ne (2,2)
\end{align*}
and the remaining four
\begin{align*} 
|\int_\Omega \cK_{\alpha\beta 1}(x) \ dx| \le 
\ve\nu\|\nabla \om\|_2^2 +c \ve^{-3/5}(\mu\lambda)^{4/5}\nu^{1/5}\|\om\|_2^{4/5} \|\om\|_2^{2} \;.
\end{align*}

For the remainder of the paper we take $\bfE=\|\omega\|_2^2$ and $\bfP=\|\nabla \omega\|_2^2$ as enstrophy and palinstrophy
respectively.  Assuming that \eqref{betacond}, \eqref{bdryint} hold, we then have that
\begin{equation}\label{Eeqn0}
\begin{aligned}
\frac{1}{2}\frac{d \bfE}{dt} + \nu\bfP \le (2\ve+\delta)\nu\bfP &+ c_2\left[\nu\lambda(\mu+\psi)
+ \frac{\nu^{1/5}}{\ve^{3/5}}(\mu\lambda)^{4/5}\bfE^{2/5}+\frac{\lambda^{1/2}}{\varepsilon\nu}\bfE\right]\bfE \\
&+\|\curl F\|_2\bfE^{1/2}\;.
\end{aligned}
\end{equation}
The point in applying the Poincar\'e inequality in \eqref{poincuse} 
is that otherwise the summand $\lambda^{3/4}\bfE^{1/2}$ would appear in the brackets.   To further simplify, we
note that the summand 
involving $\bfE^{2/5}$ will dominate both the first summand in brackets as well as the term from the force
provided
\begin{align} \label{extras}
\bfE \ge \bfE_{\text{min}}=\max\left\{\frac{\ve^{3/2}\lambda^{1/2}\nu^2}{\mu^2}(\mu+ \|\psi\|_{\infty})^{5/2},
\frac{\ve^{2/3}}{c_2^{10/9}(\mu\lambda)^{8/9}\nu^{2/9}}\|\curl F\|_2^{10/9}\right\}\;,
\end{align}
where we have tacitly assumed that $\psi \in L^{\infty}(0,T)$.
Now choose $\ve$ small enough to satisfy 
\begin{align}\label{epsdel}
2\ve+\delta =\rho < 1;.
\end{align}

Note that 
\begin{align*}
\bfE &= \|\omega\|_2^2   \le 2 \|A^{1/2}u\|_2^2  \le  2 \|u\|_2  \|A u\|_2 =2 \|u\|_2  \|\triangle u\|_2\\
    &= 2 \|u\|_2  \|\curl \omega\|_2    \le 2 \|u\|_2 \sqrt{2} \|\nabla \omega\|_2 = 2^{3/2} \bfe^{1/2}  \bfP^{1/2}
\end{align*}
hence
\begin{align}\label{PeE}
-8\bfP \le -\bfE^2/\bfe\;.
\end{align}
Gathering terms, we find that 
\begin{align}\label{Eeqn}
\frac{d \bfE}{dt} \le \left(\frac{c_2\lambda^{1/2}}{\ve\nu} -\frac{\nu(1-\rho)}{4\bfe}\right)\bfE^2 +\frac{{6}c_2(\mu\lambda)^{4/5}\nu^{1/5}}{\ve^{3/5}}\bfE^{7/5}\;.
\end{align}

We now assume, in addition to \eqref{epsdel}, that
\begin{align}\label{parcond}
\nu\bfE \ge 4\|f\|_2\bfe^{1/2}\;.
\end{align}
Using this in \eqref{energy1}, along with $\bfE \le 2\|A^{1/2}u\|_2^2$ we find that 
\begin{align}\label{Keqn}
\frac{d \bfe}{dt} \le -\frac{\nu}{2}\bfE\;.
\end{align}

\begin{figure}[htb]
\psfrag{Emax}{\tiny$\bfE_{\text{max}}$}
\psfrag{A}{\tiny(a)}
\psfrag{B}{\tiny(b)}
\psfrag{K0}{\tiny${\bfe_0}$}
\psfrag{e0}{\tiny${\bfe_0}$}
\psfrag{phi1}{\tiny$\varphi_1$}
\psfrag{phi2}{\tiny$\varphi_2$}
\psfrag{K}{\tiny${\bfe}$}
\psfrag{E}{\tiny${\bfE}$}
\psfrag{RK}{\tiny RK}
\psfrag{20}{\tiny$N=20$}
\psfrag{40}{\tiny$N=40$}
\psfrag{dKdt}{\tiny (5.10)}
\psfrag{dEdt}{\tiny (5.8)}
\psfrag{par}{\tiny (5.9)}
\psfrag{poinc}{\tiny Poincar\'e}
\psfrag{dKdt2}{\tiny (5.18)}
\psfrag{dEdt2}{\tiny (5.8)}
\psfrag{asymp}{\tiny (5.14)}
\psfrag{Emin}{\tiny (5.5)}
\psfrag{soln}{\tiny $\bfE=\varphi(\bfe)$}
\psfrag{par}{\tiny (5.9)}
  \centerline{\includegraphics[scale=.5]{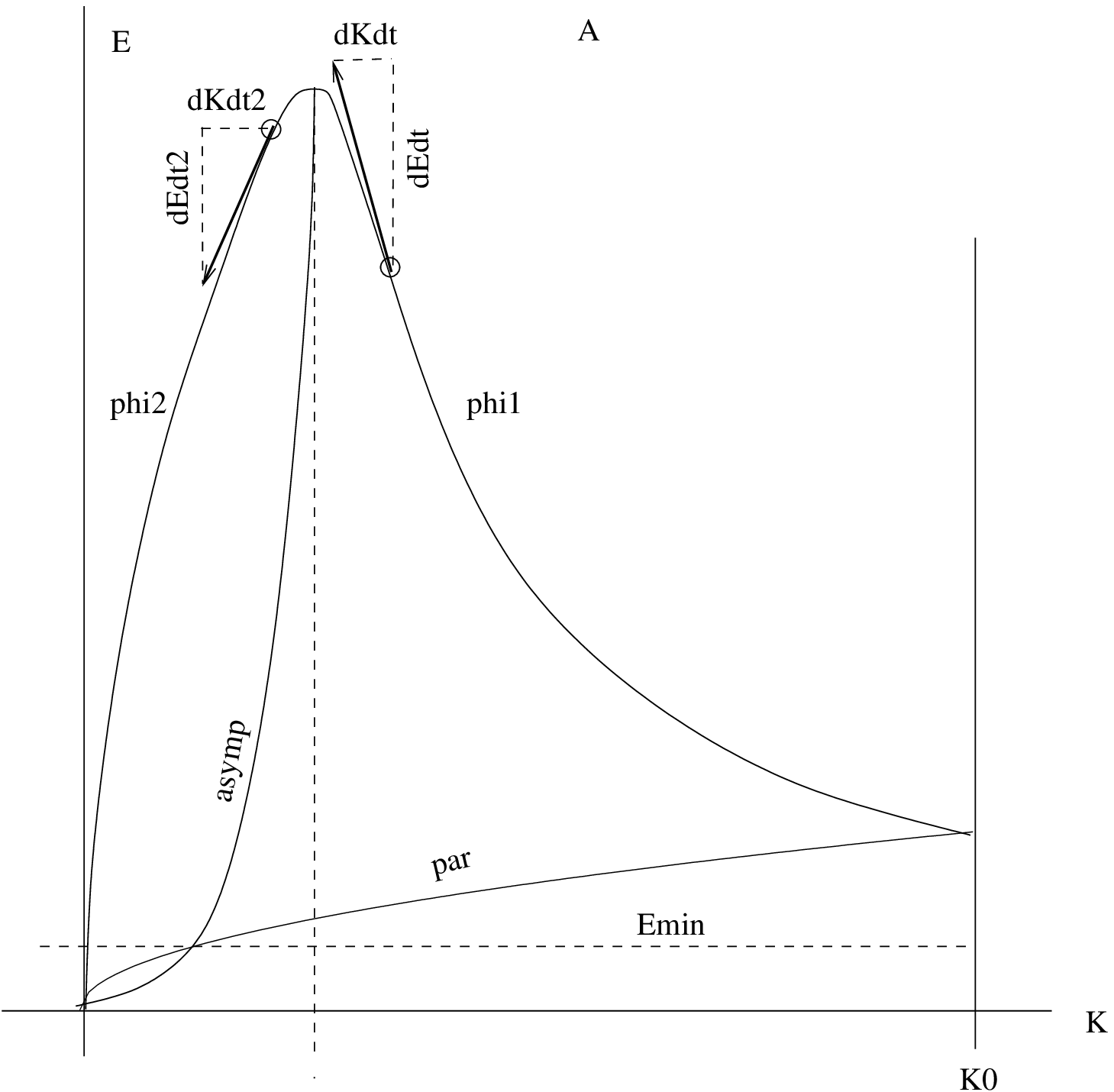}\quad \includegraphics[scale=.5]{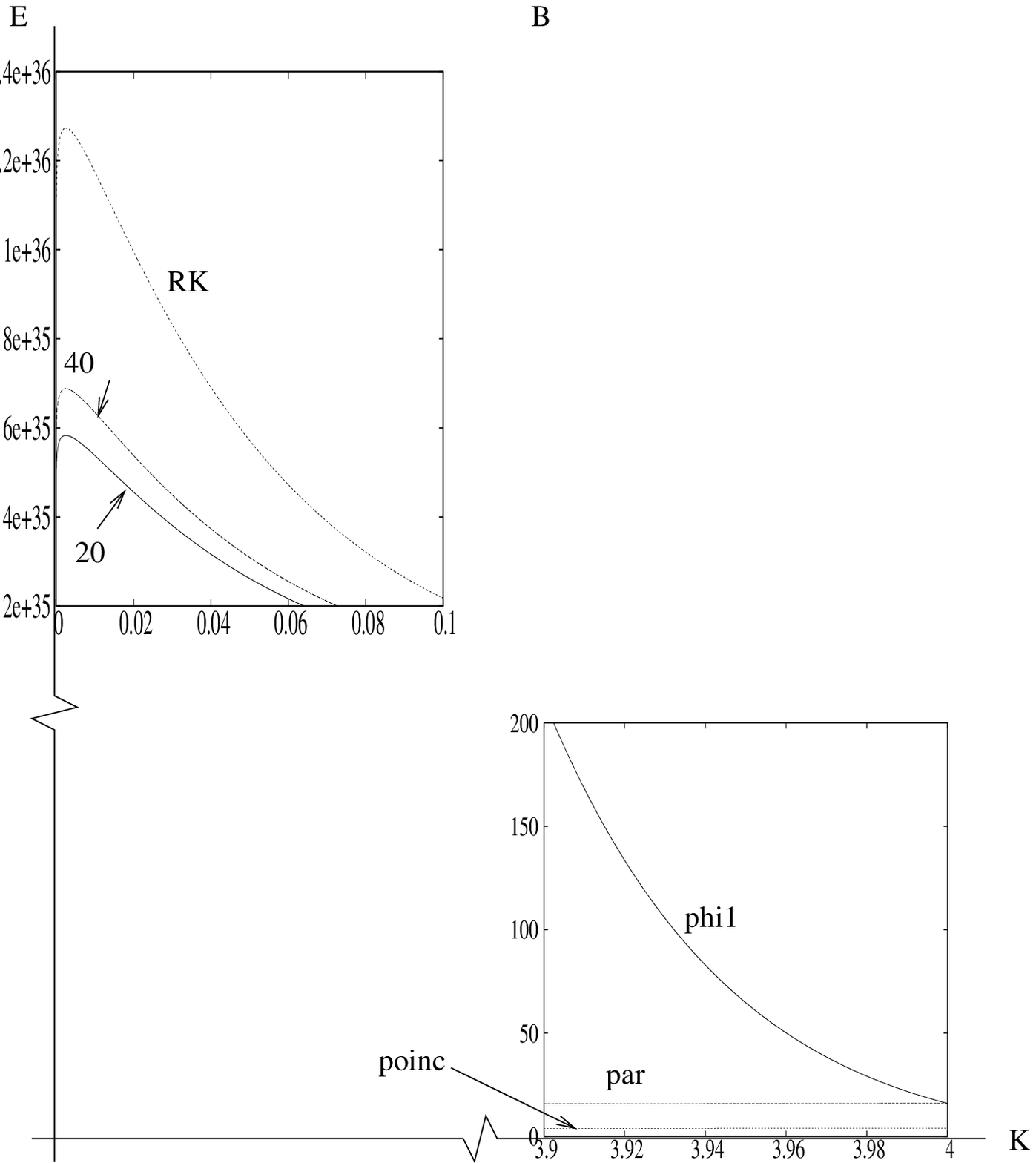}}
\caption{(a) Schematic plot indicating which bounds on time derivatives are used to determine the ODEs for $\varphi_1$
and $\varphi_2$, (b) Numerically generated bounding curve
for $\nu=\lambda=\mu=1$, $\ve=.2$, $\delta=.5$, $c_2=G=2$, by Runge-Kutta (RK) and by truncating the series approximation for the incomplete gamma function to $N$ terms.  Estimates for $\bfE_{\text{max}}$ are in Section \ref{maxest}.
}
\label{Fig2}
\end{figure}

As in Section \ref{full}, we assume (\ref{w_attr_bd}) holds, and so, by combining \eqref{Keqn} with \eqref{Eeqn} as indicated in Figure \ref{Fig2}(a)
we can construct a bounding curve for $\mathcal{A}_w$ by solving
\begin{align}\label{dEdK2}
&\frac{d \bfE}{d\bfe} =\left(\frac{1-\rho}{2\bfe}-\frac{{2}c_2\lambda^{1/2}}{\ve \nu^2}\right)\bfE - \frac{{12}c_2(\mu\lambda)^{4/5}}{\ve^{3/5}\nu^{4/5}}\bfE^{2/5}\\ 
&\bfE(\bfe_0) = \bfE_0,\quad\mbox{where} \quad \bfe_0:=\frac{\nu^2 G^2}{\lambda^{1/2}}\quad \mbox{and}\label{dEdK2_IC} \\
&\bfE_0:=\max \left\{ 4\|f\|_2\,\bfe_0^{1/2},\,\bfE_{\text{min}}\right\}=  \max\left\{4\nu^2\lambda^{1/2}G^2,\, \bfE_{\text{min}}\right\}
\end{align}
provided the right hand side in \eqref{Eeqn} is nonnegative, i.e, $(\bfe,\bfE)$ is to the right of the curve
\begin{equation}\label{Psi_curve}
\bfE=\ve^{2/3}\mu^{4/3}\lambda^{1/2}\nu^2\left[\frac{{6}\,\bfe}{\bfe_{\mathrm{a}}-\bfe}\right]^{5/3},\quad \bfe<\bfe_{\mathrm{a}}=\frac{\ve(1-\rho)}{{4}c_2}\frac{\nu^{2}}{\lambda^{1/2}}\,.
\end{equation}
Notice that if $G$ is big enough, $\bfe_0=\nu G^2/\lambda^{1/2}>\bfe_a$, and so 
the solution $\bfE=\varphi_1(\bfe)$ to (\ref{dEdK2}-\ref{dEdK2_IC}) is indeed to the right of the curve in \eqref{Psi_curve} over an interval
$(\bfe_{\max},\bfe_0]$. In fact, $\varphi_1$ is a decreasing curve on $[\bfe_{\max},\bfe_0]$  with global maximum at $(\bfe_{\max},\bfE_{\max})$ -- the intersection with the curve in \eqref{Psi_curve}.

Recognizing \eqref{dEdK2} as a Bernoulli equation, we set 
\[
\xi=\bfE^{3/5}\;, \quad a=\frac{3(1-\rho)}{10}\;, \quad b=\frac{{6}c_2\lambda^{1/2}}{5\ve\nu^2}\;, \quad \text{and } C=\frac{{36} c_2(\mu\lambda)^{4/5}}{5\ve^{3/5}\nu^{4/5}}
\]
 to obtain
\begin{align}\label{dxidK}
\frac{d \xi}{d\bfe} + \left(-\frac{a}{\bfe}+b\right)\xi = -C\;, 
\quad \xi(\bfe_0) = \xi_0= (2\nu^{3}\lambda^{1/2}G^2)^{3/5}\;.
\end{align}  
whose solution is
\begin{align}\label{xidef}
\xi(\bfe)=\bfe^{a}e^{-b\bfe}\left[\bfe_0^{-a}e^{b\bfe_0}\xi_0
-C\int_{\bfe_0}^{\bfe} s^{-a} e^{bs} \ d s \right]\;,
\end{align}

To evaluate the integral we use an expansion of the incomplete gamma function 
\begin{align}\label{series}
\gamma(\alpha,z)= \int_0^z \tau^{\alpha-1} e^{-\tau} \ d \tau=z^\alpha \sum_{n=0}^{\infty} \frac{(-z)^n}{n!(\alpha+n)}\;,
\end{align}
so that with $\alpha=1-a$
\begin{align*}
\int_{\bfe0}^{\bfe} s^{-a} e^{b s} \ d s 
&=(-b)^{a-1}\int_{-b\bfe_0}^{-b\bfe} \tau^{-a} e^{-\tau} \ d\tau \\
&=(-b)^{-\alpha}\left[\gamma(\alpha,-b\bfe)-\gamma(\alpha,-b\bfe_0)\right] \;.\\
\end{align*}
The solution to \eqref{dEdK2} can thus be written as
\begin{align*}
\bfE=\varphi_1(\bfe)
 =\left\{\bfe^{a}e^{-b\bfe}\left[ \bfe_0^{-a}e^{b\bfe_0}\xi_0
-C\left(\bfe^{\alpha} \sum_{n=0}^{\infty} \frac{(b\bfe)^n}{n!(\alpha+n)} -\bfe_0^\alpha \sum_{n=0}^{\infty} \frac{(b\bfe_0)^n}{n!(\alpha+n)}\right)\right]
\right\}^{5/3}\;.
\end{align*}

To find a bounding curve for $\bfe<\bfe_{\max}$ under the assumptions  (\ref{extras}) and (\ref{parcond}), observe that since  the right hand side in \eqref{Eeqn} is negative, we need a lower bound on $d\bfe/dt$.  From Theorem 1.1 in \cite{ChengShkoller} (henceforth, $\Omega$ is convex) there exists a constant
$c_{\Omega} \ge 1$ depending on the domain, such that $\bfoE \le c_\Omega \bfE$.  Using this in 
\begin{align*}
-2\|f\|_2 \|u\|_2 \le  \frac{d\|u\|^2_2}{dt}+ 2\nu \|A^{1/2}u\|_2^2 \le \frac{d\bfe}{dt}+ 2\nu c_{\Omega} \bfE \;,
\end{align*}
together with \eqref{parcond}, we may now write  
\begin{align}\label{lwrdedt}
-C_{\Omega}\frac{\nu}{2} \bfE \le \frac{d\bfe}{dt},\quad C_{\Omega}=1+4c_{\Omega}  \;,
\end{align}
and combine with  \eqref{Eeqn} to arrive at an initial-value problem similar to (\ref{dEdK2}-\ref{dEdK2_IC}), but with the right-hand side of \eqref{dEdK2} divided by $C_{\Omega}$, and the initial condition 
\[\bfE(\bfe_{\max})=\bfE_{\max}:=\varphi_1(\bfe_{\max}).\]
Consequently we obtain a bounding curve $\bfE=\varphi_2(\bfe)=\xi^{5/3}(\bfe)$,
where $\xi$ is as in \eqref{xidef}, but with $a$, $b$, $C$, $\bfe_0$ and $\xi_0$ replaced by
$a/C_{\Omega}$, $b/C_{\Omega}$, $C/C_{\Omega}$, $\bfe_{\max}$ and $\bfE_{\max}^{3/5}$.

It is easy to see  that $\varphi_2$ is an increasing, concave curve on $(0,\bfe_{\max})$, and that  $\varphi_2(\bfe)\searrow 0$ as $\bfe \searrow 0$. It follows that on $(0,\bfe_{\max}]$ the curve $\bfE=\varphi_2(\bfe)$ will stay to the left of that in \eqref{Psi_curve}; there will exist $\bfe_{\min}\in(0,\bfe_{\max})$ for such that $\varphi_2(\bfe_{\min})=\bfE_{\min}$ and $\varphi_2(\bfe)>\bfE_{\min}$ for all $\bfe\in(\bfe_{\min},\bfe_{\max}]$. 

In order to show that the assumption (\ref{parcond}) is satisfied on $[\bfe_{\min},\bfe_{\max}]$, note that from (\ref{xidef}), with the constants modified by $C_{\Omega}$ and with arbitrary $\bfe_0$, $0<\bfe\le\bfe_0$,
\[\xi(\bfe)\ge\bfe^{a}\frac{\xi(\bfe_0)}{\bfe_0^a},\] 
Thus, if we start above the parabola (\ref{parcond}), i.e. with $\bfE_0=\xi_0^{5/3}>4\|f\|_2\bfe_0^{1/2}/\nu$, and to the left of the curve in \eqref{Psi_curve},  then the corresponding curve $\varphi_2$ will satisfy
\[\nu\bfE\ge\nu \bfE_0\left(\frac{\bfe}{\bfe_0}\right)^\frac{1-\rho}{2C_\Omega}>4\|f\|_2\,\bfe_0^\frac{1}{2}\left(\frac{\bfe}{\bfe_0}\right)^\frac{1-\rho}{2C_\Omega}=
4\|f\|_2\,\bfe^\frac{1}{2}\left(\frac{\bfe_0}{\bfe}\right)^{\frac{1}{2}-\frac{1-\rho}{2C_\Omega}}\ge4\|f\|_2\,\bfe^\frac{1}{2},\]
and consequently, the curve $\bfE=\varphi_2(\bfe)$ will satisfy (\ref{parcond}) for $\bfe\in[0,\bfe_{\max}]$.
Therefore, $\bfE=\varphi_2(\bfe)$ will be a bounding curve for $\mathcal{A}_{w}$ for $\bfe\in[\bfe_{\min},\bfe_{\max}]$.

On $\bfe\in[0,\bfe_{\min}]$, the assumption (\ref{extras}) does not hold, and thus the construction of bounding curves depends on whether $\nu\lambda(\mu+\psi)\bfE$ dominates $\|\curl F\|_2\,\bfE^{1/2}$ in (\ref{Eeqn0}). This potentially would introduce two more cases. However, in the case $\|\curl F\|_2$ is big enough, 
e.g. \begin{equation}\label{curlF}\|\curl F\|_2\ge {c_2}\varepsilon^{3/4}\frac{\lambda^{5/4}\nu^2}{\mu}(\mu+\|\psi\|_{\infty})^{9/4},\end{equation}
the condition $\bfE\le\bfE_{\max}$ would yield $\|\curl F\|_2\bfE^{1/2}$ as the dominating term, and so instead of (\ref{Eeqn}) we will obtain
\begin{align}\label{Eeqn1}
\frac{d \bfE}{dt} \le \left(\frac{c_2\lambda^{1/2}}{\ve\nu} -\frac{\nu(1-\rho)}{4\bfe}\right)\bfE^2 +{6}\|\curl F\|_2\,\bfE^{1/2}\;,
\end{align}
which together with (\ref{lwrdedt}) yields the following equations for the bounding curve on $[0,\bfe_{\min}]$:\begin{align}\label{phi}
&\frac{d \bfE}{d\bfe} =\left(\frac{1-\rho}{2C_{\Omega}\bfe}-\frac{2c_2\lambda^{1/2}}{C_{\Omega}\ve \nu^2}\right)\bfE - \frac{{12}}{\nu C_{\Omega}} \|\curl F\|_2\,\bfE^{-1/2}\\ 
&\bfE(\bfe_{\min}) = \varphi_2(\bfe_{\min})=\bfE_{\min}\,.
\label{phi3_IC}\end{align}

Solving this Bernoulli-type equation we obtain
\[
x(\bfe)=x_0\left(\frac{\bfe}{\bfe_{\min}}\right)^{\alpha}\exp\left(\beta(\bfe_{\min}-\bfe)\right)+\gamma\int\limits_{\bfe}^{\bfe_{\min}}\left(\frac{\bfe}{t}\right)^{\alpha}\exp\left(\beta(t-\bfe)\right)\,dt\,,
\]
\[x=\bfE^{3/2},\quad x_0=(\bfE(\bfe_{\min}))^{3/2},\quad\alpha=\frac{3}{2}\frac{1-\rho}{2C_{\Omega}},\quad \beta=\frac{3c_2\lambda^{1/2}}{C_{\Omega}\ve \nu^2},\quad\gamma=\frac{{18}}{\nu C_{\Omega}} \|\curl F\|_2\,.\] 

Note that the above equations imply
\[x(\bfe)\ge x_0\left(\frac{\bfe}{\bfe_{\min}}\right)^{\alpha}\,\]
and therefore, since (\ref{parcond}) is satisfied by the initial condition (\ref{phi3_IC}), meaning 
$$x_0^{2/3}=\bfE(\bfe_{\min})>4\|f\|_2\bfe_{\min}^{1/2}/\nu,$$
we have for $\bfe\in[0,\bfe_{\min}]$
\[\bfE=(x(\bfe))^{2/3}\ge x_0^{2/3}\left(\frac{\bfe}{\bfe_{\min}}\right)^{\frac{2}{3}\alpha}=\frac{x_0^{2/3}}{\bfe_{\min}^{1/2}}\bfe^{1/2}\left(\frac{\bfe_{\min}}{\bfe}\right)^{\frac{1}{2}-\frac{2}{3}\alpha}\ge 4\|f\|_2\bfe^{1/2}/\nu\;,\]
i.e. the assumption (\ref{parcond}) holds on $\bfe\in[0,\bfe_{\min}]$, which also means that $\bfE$ will stay to the left of the curve in \eqref{Psi_curve}. Therefore, 
\[\bfE=\varphi_3(\bfe)=(x(\bfe))^{2/3}\]
is a bounding curve for $\mathcal{A}_w$ for $\bfe\in(0,\bfe_{\min}]$. It is easy to see that $\varphi_3$ is also increasing, concave and approaches zero as $\bfe\searrow0$. In fact,
\begin{equation}\label{varphi3_rate}\varphi_3(\bfe)=\cO\left(\bfe^{\frac{1-\rho}{2C_{\Omega}}}\right)\quad \mbox{as}\ \bfe\searrow 0\,.\end{equation}

We denote the complete bounding curve by
\[
\varphi(\bfe)=\begin{cases} \varphi_1(\bfe), & \bfe \in [\bfe_{\max},\bfe_0] \\
 \varphi_2(\bfe), & \bfe \in [\bfe_{\min},\bfe_{\max})\\ 
  \varphi_3(\bfe), & \bfe \in (0,\bfe_{\min})
 \;.
 \end{cases}
 \]

Since by Poincar\'e inequality $\lambda_0\bfe\le\bfoE$, and by \cite{ChengShkoller}  $\bfoE\le c_{\Omega}\bfE$, we have the following bounding region for $\mathcal{A}_w$ in $(\bfe,\bfE)$-plane:
\[
\underline{\lambda}\,\bfe\,\le\,\bfE\,\le\,\phi(\bfe)\,. \]
with $\underline{\lambda}=\lambda_0/c_{\Omega}$.

Numerical solutions of the bounding curve $\varphi$ are plotted in Figure 2(b).  One is made by using a 4th order Runge-Kutta method to solve
the ODE, the other two by truncating the series approximation for the incomplete gamma function.  Though the Grashof number
is quite small ($G=2$), the maximum value $\bfE_{\text{max}}$ is at least $\cO(10^{35})$.  Rigorous upper and lower bounds
for $\bfE_{\text{max}}$  in the next section confirm the dramatic rise in this bounding curve.

In particular, estimates of Section  \ref{maxest} show that as the Grashof number grows, $\bfE_{\max}\sim \exp(c_3 G^2)$, $\bfe_a\sim\bfe_{\max}\sim G^0$, and therefore, we can deduce from (\ref{xidef}) that $\bfe_{\min}\sim \exp(-c_4G^2)$, for appropriate $G$-independent constants $c_3$, $c_4$ 
Thus as $G\to\infty$,  
the possible big values for enstrophy will be constrained in a narrow (compared to $G$) strip of energy values located in the proximity of $\bfe=0$.

\vskip .15truein 



\section{Estimates for the maximum value of $\bfE=\varphi(\bfe)$} \label{maxest}

For simplicity we examine the case $\nu=\lambda=1$, so that
the bounding curve $\bfE=\varphi_1(\bfe)$ in the 
critical case is the solution to 
\begin{equation}\label{ivp1}
\begin{aligned}
\frac{d \bfE}{d \bfe}=-\left(\frac{2c_2}{\ve}-\frac{1-\rho}{2\bfe}\right)\,\bfE-\frac{12c_2\mu}{\ve^{3/5}}\, \bfE^{2/5},\\
\bfE(\bfe_0)=\bfE_0=4\|f\|_2\bfe_0^{1/2}=4G^2\,.
\end{aligned}
\end{equation}\subsection{A lower bound:}
\ \\

Clearly, 
\[
\frac{d \bfE}{d \bfe} {\le} -\left(\frac{2c_2}{\ve}-\frac{1-\rho}{2\bfe}\right)\,\bfE\,.
\]
Therefore, on $[\bfe_{\max},\bfe_0]$ the curve $\bfE=\varphi(\bfe)$ lies above the solution to
\begin{equation}\label{ivp2}
\begin{aligned}
\frac{d \tilde{\bfE}}{d\bfe}=-\left(\frac{2c_2}{\ve}-\frac{1-\rho}{2\bfe}\right)\,\tilde{\bfE},\\
\bfe_0=G^2,\quad \bfE(\bfe_0)=\bfE_0=4G^2\,,
\end{aligned}
\end{equation}
which  is
\[
\tilde{\bfE}(\bfe)=\bfE_0\left(\frac{\bfe}{\bfe_0}\right)^{\frac{1-\rho}{2}}\exp\left({\frac{2c_2}{\ve}(\bfe_0-\bfe)}\right)=
4G^{1+\rho}\bfe^{\frac{1-\rho}{2}} \exp\left({\frac{2c_2}{\ve}(G^2-\bfe)}\right)
\]

It is immediate from the ODE in (\ref{ivp2}) that the maximum for $\tilde{\bfE}$ is achieved when 
$$
\bfe=\bfe_{\text{crit}}=\frac{\ve(1-\rho)}{4c_2},
$$ and therefore 
the maximal value of the bounding curve $\bfE=\varphi(\bfe)$ satisfies:
\begin{equation}\label{lwr_bd}
\bfE_{\text{max}}\ge 
4 G^{1+\rho}\bfe_{\text{crit}}^{\frac{1-\rho}{2}} \exp\left(\frac{2c_2}{\ve}(G^2 - \bfe_{\text{crit}})\right)\,.
\end{equation}

\subsection{An upper bound:}
\ \\

Looking back at the original equations (\ref{ivp1}), we notice that if for $\eta>0$
\begin{equation}\label{cond_1.1}
\frac{\eta}{\ve}\,\bfE\ge\frac{12\mu}{\ve^{3/5}}\, \bfE^{2/5},\quad 
\text{or, equivalently} \quad 
\eta\ge \frac{12\mu}{\ve^{2/5}\bfE^{3/5}}\,,
\end{equation}
then
\[
\frac{d \bfE}{d \bfe}\ge-\left(\frac{2c_2}{\ve}-\frac{1-\rho}{2\bfe}\right)\,\bfE - \frac{\eta c_2}{\ve}\,\bfE=  -\left(\frac{(2+\eta)c_2}{\ve}-\frac{1-\rho}{2\bfe}\right)\,\bfE\,.
\]
Note if \eqref{cond_1.1} holds for $\bfE=\bfE_0$, then it holds for the bounding curve $\bfE=\phi(\bfe)$ over $\bfe\in[\bfe_{\text{max}}, \bfe_0]$. 
Thus, for  $\eta$ big enough so that (\ref{cond_1.1}) holds at $\bfE=\bfE_0$, on the interval $[\bfe_{\text{max}}, \bfe_0]$ the bounding curve $\bfE=\varphi(\bfe)$ will lie below the solution to the following initial value problem:
\begin{equation}\label{ivp2.1}
\begin{aligned}
\frac{d \tilde{\bfE}}{d \bfe}=-\left(\frac{(2+\eta)c_2}{\ve}-\frac{1-\rho}{2\bfe}\right)\,\tilde{\bfE},\\
\bfe_0=G^2,\quad \tilde{\bfE}(\bfe_0)=\bfE_0=4G^2\,.
\end{aligned}
\end{equation}

Obviously, this is exactly  (\ref{ivp2}) with  $2c_2$ replaced by $(2+\eta)c_2$. Therefore,
\begin{equation}\label{uppr_bd}
\bfE_{\text{max}}\le 4 G^{1+\rho}\bar{\bfe}_{\text{crit}}^{\frac{1-\rho}{2}} \exp\left(\frac{(2+\eta)c_2}{\ve}(G^2 - \bar{\bfe}_{\text{crit}})\right) \;,
\end{equation}
where
$$
\bar{\bfe}_{\text{crit}}=\frac{\ve(1-\rho)}{2(2+\eta)c_2}\;.
$$
Note that $\bar{\bfe}_{\text{crit}}\le \bfe_{\max} \le {\bfe}_{\text{crit}}$, so $\bfe_{\max} \sim \ve$. 

Taking  $\ve=.2$, $\mu=1$ $G=c_2=2$ in \eqref{cond_1.1} we can choose $\eta=4.33$ to evaluate the expressions in \eqref{lwr_bd} and \eqref{uppr_bd} and find that
$$
5.83 \cdot 10^{35} \le \bfE_{\max} \le 9.13 \cdot 10^{110},
$$
which is consistent with the plot in Figure \ref{Fig2}(b). 
Naturally, since $G$ was chosen so small, the upper bound on $E_{\max}$ is much less sharp than the lower bound. 
We can make the upper bound sharper if we set $\tilde{\bfE}(\bfe_0)=39311.12\gg 4G^2$ in (\ref{ivp2.1}). Then we can choose $\eta=0.04$ resulting in the upper bound $E_{\max}\le 6.99\cdot 10^{39}$, which is much closer to the lower bound.
Note that as $G$ increases, so does $\bfE_0$, and thus the smaller we may take $\eta$ to sharpen the upper bound. 

\section{{The sub-critical case $1/2 < r \le 1$}}\label{subcritsec}

Instead of \eqref{HolderPoint} we have 
\begin{align}\label{HolderPoint2}
\int_\Omega |\cK_{222}(x)| \ dx &\le c\lambda^{r/2} \int_\Omega |\omega(x)|^2 
\left(\int_\Omega \frac{|\omega(y)|}{|x-y|^{3-r}}\ dy\right)\ dx \\
&\le c\lambda^{r/2} \|\omega\|_{8/3}^2\| h*|\omega| \|_4 \;,
 \end{align}
 where 
 $$
 h(x)=\frac{1}{|x|^{3-r}}\;.
 $$
To estimate the convolution, we apply the
H\"older inequality for convolutions according to
$$
1+\frac{1}{4}=\frac{1}{p}+\frac{1}{q}\quad \text{where} \quad p=\frac{3}{3-r}\;, 
q=\frac{12}{3+4r}
$$
followed by interpolation to obtain
\begin{align*}
\| h*|\omega| \|_4 \le c\|h\|_{p,w} \|\omega\|_q \le c  \|\omega\|_q \le 
c\|\omega\|_2^{\frac{1+4r}{4}}\|\nabla \omega\|_2^{\frac{3-4r}{4}}\;.
\end{align*}
Using this and the interpolation estimate
$$
\|\omega\|^2_{8/3} \le c \|\omega\|_2^{5/4} \|\nabla \omega\|^{3/4}_2
$$
into \eqref{HolderPoint2},  
we have
\begin{align*}
\int_\Omega |\cK_{222}(x)| \ dx & \le c\lambda^{r/2}\|\omega\|_2^{\frac{3+2r}{2}}
\|\nabla \omega\|_2^{\frac{3-2r}{2}} \\
&\le \ve \nu \|\nabla \omega\|_2^2 + c\left(\frac{\lambda^{2r}}{(\ve\nu)^{3-2r}}\right)^{\frac{1}{1+2r}}
\|\omega\|_2^{\frac{6+4r}{1+2r}}\;.
 \end{align*}
 
This gives a bound just as in \eqref{Eeqn0}, except that 
\begin{align*}
\frac{\lambda^{1/2}}{\varepsilon\nu}\bfE \quad \text{is replaced by} \quad \left(\frac{\lambda^{2r}}{(\ve\nu)^{3-2r}}\right)^{\frac{1}{1+2r}}\bfE^{\frac{2}{1+2r}}\;.
\end{align*}
Thus, under a threshold condition 
\begin{equation}\label{underlineE}\bfE\ge\underline{\bfE},\end{equation}
where $\underline{\bfE}$ is defined similarly to $\bfE_{\min}$ in \eqref{extras} so that the summand 
with the $\frac{2}{1+2r}$ power dominates the others, we have
\begin{align}\label{subdEdt_0}
\frac{d\bfE}{dt} \le -\frac{\nu(1-\rho)}{{4}\,\bfe}\bfE^2+ \nu \frac{C}{2}\bfE^{\frac{3+2r}{1+2r}}
\end{align} 
where 
$$
C=\frac{2}{\nu}\,{6c}\left(\frac{\lambda^{2r}}{(\ve\nu)^{3-2r}}\right)^{\frac{1}{1+2r}}\;.
$$
We then combine \eqref{subdEdt} with \eqref{Keqn} to find a bounding curve solving
\begin{align}\label{subdEdt}
\frac{d\bfE}{d\bfe} = \frac{1-\rho}{2\,\bfe}\bfE- C\bfE^{\frac{2}{1+2r}}
\end{align} 
over the interval $[\bar \bfe, \bfe_0]$ where \eqref{subdEdt_0} is nonnegative.  The solution to
\eqref{subdEdt} is 
\begin{align}\label{subbed}
\bfE=\phi_1(\bfe)=\left\{\left(\frac{\bfe}{\bfe_0}\right)^{\alpha\sigma}\bfE_0^\sigma +\frac{\sigma C}{1-{\alpha}\sigma}\left[
\frac{\bfe^{{\alpha}\sigma}}{\bfe_0^{{\alpha}\sigma-1}}-\bfe\right]\right\}^{1/\sigma}\;,
\end{align}
where 
$$
\sigma= \frac{2r-1}{1+2r}, \quad\mbox{and}\quad \alpha=\frac{1-\rho}{2}\;.
$$
From \eqref{subdEdt}, we have that the maximum value, $\bar \bfE=\phi_1(\bar{\bfe})$ is achieved where the bounding curve (\ref{subbed}) intersects the curve $\alpha\bfE^\sigma=C\bfe$.   An elementary calculation
gives, for appropriate $G$-independent constants $C_1$ and $C_2$,
$$
\bar{\bfE}=\left[C_1(G^2)^{\sigma-\alpha\sigma}+C_2(G^2)^{1-\alpha\sigma}
\right]^{\frac{1}{\sigma(1-{\alpha}\sigma)}}
=\cO(G^{2/\sigma})
$$
as  $G \to \infty$.

For $\bfe < \bar{\bfe}$, as long as (\ref{underlineE}) and (\ref{parcond}) hold, we combine (\ref{subdEdt_0}) with \eqref{lwrdedt} to obtain a bounding curve 
$\bfE=\phi_2(\bfe)$ where $\phi_2$ is the same as $\phi$ in \eqref{subbed} except ${\alpha}$, $C$, $\bfe_0$ and $\bfE_0$ 
are replaced by $\alpha/C_{\Omega}$, $C/C_{\Omega}$, $\bar \bfe$ and $\bar \bfE$.
As in Section \ref{critic}, we can show that $\phi_2$ is an increasing concave curve that converges to zero at the origin and stays above the parabola $\bfE=4\|f\|_2\,\bfe^{1/2}/\nu$ over the interval $[0,\bar{\bfe}]$. 
Therefore $\phi_2$ will intersect the line $\bfE=\underline{\bfE}$ at some point 
$\underline{\bfe}\in(0,\bar{\bfe})$, and so 
\[\bfE=\phi_2(\bfe)\]
is a bounding curve for $\mathcal{A}_w$ on $[\underline{\bfe},\bar{\bfe}]$.

The bounding curve for $0 < \bfe < \underline{\bfe}$ is obtained analogously to $\varphi_3$ in Section \ref{critic}. In fact if $\bfE\le\underline\bfE$ and $\|\curl F\|_2$ is bigger than an appropriate Grashof-independent threshold similar to (\ref{curlF}),
then the term $\|\curl F\|_2\,\bfE^{1/2}$ will dominate, resulting in the following equivalent of (\ref{phi}-\ref{phi3_IC}):
\begin{align}\label{phi-s}
&\frac{d \bfE}{d\bfe} =\frac{1-\rho}{2C_{\Omega}\bfe}\,\bfE - \frac{{12}}{C_{\Omega}} \|\curl F\|_2\,\bfE^{-1/2}\\ 
&\bfE(\underline{\bfe}) = \phi_2(\underline{\bfe})\,.
\label{phi3_IC-}\end{align}
This yields the curve
\begin{equation}
\bfE=\phi_3(\bfe):=\left\{\left(\frac{\bfe}{\underline{\bfe}}\right)^{3\beta/2}\underline{\bfE}^{3/2} +\frac{{36}\|\curl F\|_2}{C_{\Omega}(2-3\beta)}\left[
\frac{\bfe^{3\beta/2}}{\underline{\bfe}^{3\beta/2-1}}-\bfe\right]\right\}^{2/3}
\end{equation}
with $\beta=(1-\rho)/(2C_{\Omega})$. As in Section \ref{critic}, $\phi_3(\bfe)$ is increasing on $(0,\underline{\bfe}]$, convergent to zero at $0$, and is above the parabola $\bfE=4\|f\|_2\,\bfe^{1/2}/\nu$. Also,
\begin{equation}\phi_3(\bfe)=\cO\left(\bfe^{\frac{1-\rho}{2C_{\Omega}}}\right)\quad \mbox{as}\ \bfe\searrow 0\,,\end{equation}
i.e. {\em the same rate} as in critical case -- see (\ref{varphi3_rate}). 

Thus we obtain the following bounding region for $\mathcal{A}_w$ in $(\bfe,\bfE)$-plane:
\[
\underline{\lambda}\,\bfe\,\le\,\bfE\,\le\,\phi(\bfe):=\begin{cases} \phi_1(\bfe), & \bfe \in [\bar{\bfe},\bfe_0] \\
 \phi_2(\bfe), & \bfe \in [\underline{\bfe},\bar{\bfe})\\ 
  \phi_3(\bfe), & \bfe \in (0,\underline{\bfe})
 \;.
 \end{cases}
 \]

\begin{figure}[h]
       \psfrag{e}{{\tiny${\bfe}$}}
      \psfrag{E}{{\tiny ${\bfE}$}}
            \psfrag{I}{\tiny I}     
      \psfrag{II}{\tiny II}     
      \psfrag{III}{\tiny III}    
   \centerline{\includegraphics[scale=.4]{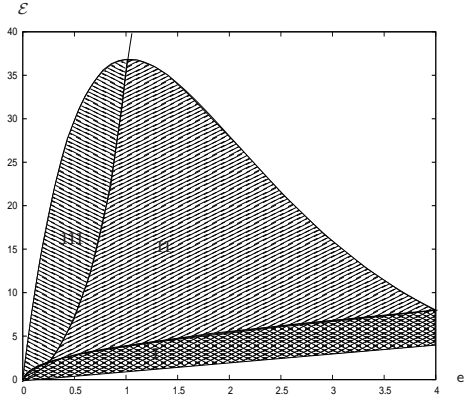}}
\caption{Bounding curves for subcritical case, $r=.51$, $G=2$, $C=1$. Region I is recurrent.  In region III the enstrophy is always decreasing, while
in regions II and III the energy is always decreasing. }
\label{fig3}
\end{figure}

These bounds are illustrated in Figure \ref{fig3}.

\section{{Bounds from scaling-invariant regularity condition}}\label{scaling-invariant}

A natural question to pose at this point--for the sake
of comparison--is what are 
the effects of the classical, analytical
\emph{scaling-invariant} regularity criteria on the vorticity magnitude on the dynamics 
considered in the energy-enstrophy plane. As an illustration, 
we present the case of the small-$L^\frac{3}{2}$ spatial integrability 
condition,
\[
 \sup_{t \in (0,T)} \|\omega(t)\|_\frac{3}{2} \le \epsilon_0,
\]
for a sufficiently small constant $\epsilon_0$.

In this case, the estimate on the vortex-stretching term reads as follows.

\begin{align*}
\int_\Omega |\omega \cdot \nabla u \cdot \omega| \, dx
                       &\le \|\omega\|_\frac{3}{2} \|\nabla u\|_6 \|\omega\|_6\\
                       &\le C_\Omega \, \|\omega\|_\frac{3}{2} \|\omega\|_6^2
                       \le c_\Omega \, \|\omega\|_\frac{3}{2} \biggl( \|\nabla \omega\|_2^2 
                       + k_\Omega \|\omega\|_2^2\biggr)\\
                       &\le  \epsilon_0 \, c_\Omega \, \biggl( \|\nabla \omega\|_2^2 
                       + k_\Omega \|\omega\|_2^2\biggr)\\
                       & \le  \epsilon_0 c'_\Omega\nu \left(\|\nabla \omega\|_2^2
                       +\lambda\|\omega\|_2^2\right).
\end{align*}

Using this along with \eqref{bdryint}, \eqref{PeE} yields
\begin{align*} 
\frac{d \bfE}{dt} 
& \le -\frac{\nu}{4} (1-\epsilon_0 c'_\Omega-\delta)\frac{\bfE^2}{\bfe}+2\nu\lambda(\psi+\epsilon_0 c'_\Omega) \bfE
+2\|\curl F\|_2 \bfE^{1/2} .
\end{align*}
Now assume 
\begin{align*} 
\bfE  \ge \left(\frac{\|\curl F\|_2}{\nu\lambda(\psi+\epsilon_0 c'_\Omega)}\right)^2\;,
\end{align*}
so that
\begin{align*} 
\frac{d \bfE}{dt} \le -\frac{\nu\alpha}{2\bfe}\bfE^2 + \frac{\nu \beta}{2} \bfE
\end{align*}
where if we take $\epsilon_0$, $\delta$ small enough
\begin{align*} 
0<\alpha=\frac{1}{2}(1-\epsilon_0c'_\Omega-\delta)< 1 \qquad \beta=8\lambda(\psi+ \epsilon_0c'_\Omega)\;.
\end{align*}
We combine with \eqref{Keqn}, and thus consider 
\begin{align*} 
\frac{d\bfE}{d\bfe}=\frac{\alpha}{\bfe}\bfE-\beta\;, \qquad \bfE(\bfe_0)=\bfE_0
\end{align*} 
whose solution is given by 
\begin{align*} 
\bfE=\left(\frac{\bfE_0}{\bfe_0^\alpha} + \frac{\beta\bfe_0^{1-\alpha}}{1-\alpha}\right) \bfe^\alpha-
\frac{\beta}{1-\alpha} \bfe \;.
\end{align*}
This curve provides a bound as long as $d\bfE/d\bfe < 0$, i.e., for
$$
\bfe_{\max}=\left[ \frac{\alpha(1-\alpha)}{\beta}\left(\frac{\bfE_0}{\bfe_0^\alpha} + \frac{\beta\bfe_0^{1-\alpha}}{1-\alpha}\right)\right]^{\frac{1}{1-\alpha}}\ < \bfe \le \bfe_0;,
$$
which gives a maximal value of
$$
\bfE_{\max}=\bfE(\bfe_{\max})\sim G^{4-2\alpha}
$$
(ignoring the dependence on $\nu$, $\lambda$).

This is somewhat unexpected. With respect to the 
scaling associated with the 3D NSE, the $L^\frac{3}{2}$-condition is scaling-invariant, while the 
$\frac{1}{2}$-H\"older coherence condition exhibits sub-critical scaling.
Hence, from this point of view, the scaling-invariant condition is
a weaker condition. In contrast, when considering dynamics of 
the solutions, at least within the realm of bounding curves in the energy,enstrophy 
plane, the $\frac{1}{2}$-H\"older 
coherence condition appears to be much weaker as it allows for \emph{qualitatively} higher 
maximum values of the enstrophy: exponential \emph{vs.} algebraic (in Grashof). 
Incidentally, this is philosophically consistent with another very recent result
illustrating discrepancies between the scaling and dynamical properties of solutions to the
3D NSE \cite{vignette}.

To compare the $L^\frac{3}{2}$ scaling-invariant case to the $r$-H\"older coherence assumption
in the case $\frac{1}{2} < r \le 1$, in terms of $G$,
we set 
\begin{align*}
\frac{4r+2}{2r-1}=4-2\alpha=3+\epsilon_0c_\Omega' +\delta\;, \quad \text{i.e.,} \quad
r=\frac{\epsilon_0c_\Omega' +\delta +5}{  2(\epsilon_0c_\Omega' +\delta)+2}
\end{align*}
and find that 
$$
0 < \epsilon_0c_\Omega' +\delta < 1 \implies 3/2 < r < 5/2\;.
$$
Since 
\begin{align*}
\frac{4r+2}{2r-1}>4-2\alpha \quad \text{for} \quad 1/2 < r < 3/2,
\end{align*}
we conclude that even
the stronger $\frac{1}{2} < r \le 1$ H\"older coherence assumption--in principle--allows for larger values of
the maximal enstrophy than the $L^\frac{3}{2}$-integrability condition, and is in this sense
a dynamically weaker condition.

\bibliographystyle{plain}
\bibliography{dfgj}

\end{document}